\documentclass[aps,prl,preprint,superscriptaddress, longbibliography]{revtex4-2}
\usepackage{enumitem}
\usepackage{amsmath}
\usepackage{slashed} 
\usepackage{amsfonts,color}
\usepackage{hyperref}

\begin{document}


\title{An axion-like mechanism  for confinement in QCD}


\author{J. L. Alonso}
\affiliation{Departamento de F\'{\i}sica Te\'orica, Universidad de Zaragoza,  Campus San Francisco, 50009 Zaragoza (Spain)}
\affiliation{Instituto de Biocomputaci{\'{o}}n y F{\'{\i}}sica de Sistemas Complejos (BIFI), Universidad de Zaragoza,  Edificio I+D, Mariano Esquillor s/n, 50018 Zaragoza (Spain)}
\affiliation{Centro de Astropart{\'{\i}}culas y F{\'{\i}}sica de Altas Energías (CAPA),
Departamento de F{\'{\i}}sica Te\'orica, Universidad de Zaragoza, Zaragoza 50009, (Spain)}
\author{C. Bouthelier-Madre}
\affiliation{Departamento de F\'{\i}sica Te\'orica, Universidad de Zaragoza,  Campus San Francisco, 50009 Zaragoza (Spain)}
\affiliation{Instituto de Biocomputaci{\'{o}}n y F{\'{\i}}sica de Sistemas Complejos (BIFI), Universidad de Zaragoza,  Edificio I+D, Mariano Esquillor s/n, 50018 Zaragoza (Spain)}

\affiliation{Centro de Astropart{\'{\i}}culas y F{\'{\i}}sica de Altas Energías (CAPA),
Departamento de F{\'{\i}}sica Te\'orica, Universidad de Zaragoza, Zaragoza 50009, (Spain)} 
\author{J. Clemente-Gallardo}
\affiliation{Departamento de F\'{\i}sica Te\'orica, Universidad de Zaragoza,  Campus San Francisco, 50009 Zaragoza (Spain)}

\affiliation{Instituto de Biocomputaci{\'{o}}n y F{\'{\i}}sica de Sistemas Complejos (BIFI), Universidad de Zaragoza,  Edificio I+D, Mariano Esquillor s/n, 50018 Zaragoza (Spain)}
\affiliation{Centro de Astropart{\'{\i}}culas y F{\'{\i}}sica de Altas Energías (CAPA),
Departamento de F{\'{\i}}sica Te\'orica, Universidad de Zaragoza, Zaragoza 50009, (Spain)}

\author{D. Martínez-Crespo}
\affiliation{Departamento de F\'{\i}sica Te\'orica, Universidad de Zaragoza,  Campus San Francisco, 50009 Zaragoza (Spain)}
\affiliation{Centro de Astropart{\'{\i}}culas y F{\'{\i}}sica de Altas Energías (CAPA),
Departamento de F{\'{\i}}sica Te\'orica, Universidad de Zaragoza, Zaragoza 50009, (Spain)} 
%


\date{\today}

\begin{abstract}
    This {letter} is about confinement in QCD.  At the moment we have pictures of confinement  to complete our understanding of the physics of strongly interacting particles, interaction which asks for it. As it is said in \cite{Burgess2006} : 
    `` In principle it should be possible to derive the conﬁnement hypothesis  from the QCD Lagrangian. At this time, no rigorous derivation exists, 
    so it is not absolutely clear that the conﬁnement hypothesis is a \textit{bone 
    ﬁde} prediction of QCD".
    In this letter we show that a sufficient (of course not necessary) condition for confinement is  that  topological structure of vacuum in Nature does not correspond to the $\theta$--vacuum. Therefore, if different vacua with nontrivial winding number cannot be efficiently connected by tunneling,  we obtain confinement  as a consequence. Moreover, we show that an axion-like mechanism would suffice to describe a confined QCD phase. 
\end{abstract}


\maketitle

\section{Introduction}


 In quantum chromodynamics (QCD), the fact of color charged particles not being directly observed is justified with the phenomenon called \textit{confinement}, which  forbids the existence of isolated color charged particles. Thus because of confinement,  quarks and gluons must appear combined in the form of hadrons to be stable. Despite the simplicity and frequent use of the assertion it is a difficult task to obtain confinement as a direct consequence of QCD postulates. The aim of this letter is to provide a sufficient condition for existence of confinement as a condition on the structure of QCD $\theta$-vacuum and the nature of the gauge group. From this point of view (the structure of the $\theta$ vacuum), we will begin by considering the role of instantons, even if similar conclusions can be obtained from the topological structure of the gauge group, as we will briefly mention below.


The original motivation to study the role of instantons  in QCD has been to describe the transition from weak to strong coupling (see \cite{Callan1979,Luscher1978,Callan1978}). Instantons were not aimed to solve color confinement.  Instead, the project was, describing strong interactions within QCD and assuming confinement, to show that everything which was known about QCD was consistent with the notion that instantons bridge the gap between weak-and strong-coupled physics \cite{Callan1978}. Now we can say that this project has been successful. Indeed, now
it is a fact that instantons, on the one hand, and pictures of confinement (or confinement criteria), on the other, have been revealed as two fundamental ingredients \cite{Deur2016,Boucaud2004} to study the coupling $\alpha_s (Q^2 )$. This function sets the strength of the interactions involving quarks and gluons in QCD, as a function of the momentum transfer $Q$, over the complete $Q^2$ range in order to describe hadronic interactions at both long and short distances.    That is, instanton effects could carry the theory all the way into the strong-coupling regime \cite{Callan1979,Luscher1978,Callan1978}.

{ Going back to the confinement in QCD, let us recall that there is not yet a proof of color confinement in any non-abelian gauge theory}. The phenomenon can be qualitatively { explained} by {assum}ing that the gluon field between a pair of color charges forms a narrow flux tube between them. From this stem the string-fragmentation models were invented to account for the fact that when quarks are produced in the particle accelerator, instead of seeing the individual quarks in detectors, many color-neutral particles are {detected}. This process is called hadronization, fragmentation, or string breaking. One of the most studied models of this type is Lund string fragmentation model \cite{Andersson1998}.

In this letter we { consider the relation between both concepts, instantons and confinement, and wonder how the screening of instanton configurations in the pure gauge case may affect {E}uclidean QCD.  We will show below how, if we modify instanton configurations in our picture, we may have a vacuum structure different from the $\theta$ vacuum of QCD. If this happens (see below) the Cluster Decomposition Property (CDP) no longer holds. In that case,  because of a theorem discussed in \cite{Lowdon2016}, and under some additional technical requirements, confinement must appear. In addition, we show how an axion-like mechanism {
{ may be sufficient} to provide a confined phase of the QCD sector.   

The structure of the letter is as follows. In the next section we recall the role of winding numbers in the vacuum of QCD. Then, in the second and third sections we present the main contributions of the letter, analyzing the relations between instantons, $\theta$-vacua, the Cluster Decomposition Property and Lowdon theorem. From that analysis we conclude, in the final section,  that if we impose conditions on the topological structure of the vacuum and the gauge group, confinement must appear. {One possible mechanism to impose these conditions is to consider the effect of the axion-like field introduced in \cite{dvali2022}, which makes quantum tunneling to be mediated only by massive modes.}

}
\section{ Winding numbers: $\theta$--vacua and instantons}
\label{sec:winding}
 Instantons are solutions of non-abelian Yang-Mills equations in Euclidean space with non-vanishing first Chern class. Let us briefly recall now the main definitions and properties of these objects and the necessity of including vacua labelled by these integer numbers.
 Our argument is essentially semiclassical in nature matching most of the usual arguments about $\theta$-vacua available in the literature. It is also possible to derive the content on this section from the topology of the gauge group \cite{Strocchi2019}, even though we will not pursue that approach in this paper.  A great account of this discussion can be found in \cite{Gomes2020}.

 \subsection{Why winding numbers?}

The QCD Lagrangian density is
    
    \begin{equation}
    \label{eq:lcolor}
    \mathcal{L}=-\frac{1}{4} \operatorname{Tr}[F^{\mu \nu} F_{\mu \nu}]+\sum_{\alpha} \bar{\psi}_{j}^{(\alpha)}\left(i \slashed{D}_{j k}-m^{(\alpha)} \delta_{j k}\right) \psi_{k}^{(\alpha)}
    \end{equation}
    where { the second term represents the fermion part} and $F^{\mu \nu}$ is the $\mathfrak{su}(3)$-valued coordinate expression of the curvature $F$ of an  Ehresmann connection on a  $SU(3)$ principle bundle $\pi:P\to \mathbb{R}^{1,3}$. In particular a single chart is enough to cover the whole Minkowskian space time and the connection is fully specified by a gauge potential $A$. 

    Let us consider for now the pure gauge sector of the model. For physical reasons, some conditions must be imposed on the gauge potential to make the action, associated with the Lagrangian $L$, the spatial integral of 
    $\mathcal{L}$, finite. In particular,  $A$ must vanish at spatial infinity $S_{\infty}^2$ while approaching a curvature free configuration  over the asymptotic past and future Cauchy hypersurfaces $\Sigma_{\pm}$.  
    In geometrical terms, this implies, see \cite{Gomes2020}, that the First integral  Chern class $\mathrm{Ch}[P]$ must be an integer  with 

    \begin{equation}
    \label{eq:cherntermint}
    \mathrm{Ch}[P]=\frac{1}{8 \pi^{2}} \int_{\mathbb{R}^{1,3}}  \operatorname{Tr}[F\wedge F].
    \end{equation}
    
    A crucial fact about this number is that it is the integral of a local operator
    \begin{equation}
    \label{eq:chernterm}
    \frac{1}{8 \pi^{2}}\operatorname{Tr}[F\wedge F]
    \end{equation}
    that can be included in the Lagrangian \eqref{eq:lcolor} { without breaking} Lorenz invariance.      Moreover, on one chart, { the form is exact and hence} $\frac{1}{8 \pi^{2}}\operatorname{Tr}[F\wedge F]=d \operatorname{cs}_A $ with $ \operatorname{cs}_A$ the Chern-Simons form
     \begin{equation}
    \label{eq:ChernSimonsDensity}
     \operatorname{cs}_A= \frac{1}{8 \pi^{2}} \operatorname{tr}\left(A \wedge \mathrm{d} A+\frac{2}{3} A \wedge A \wedge A\right).
    \end{equation}
    { This allows us to write the Chern class as an integral over the boundaries 
    \begin{equation}
    \label{eq:chwinding}
    \mathrm{Ch}[P]= \int_{\Sigma_+}cs_A-\int_{\Sigma_{-} }cs_A=n_+-n_{-},
    \end{equation}
    where} $n_\pm$ are winding numbers for field configurations on the asymptotic past and future of the theory. Because of the asymptotic behavior imposed  on the fields, the winding number can be casted in more familiar terms by means of the Wess-Zumino invariant that, over $\Sigma_{\pm}$, takes the form

    \begin{equation}
    \label{eq:windingNumber}
    n=\frac{i}{24 \pi^{2}} \int_\Sigma  \epsilon^{i j k} \operatorname{Tr}\left({A}_{i} {A}_{j} {A}_{k}\right)
    \end{equation}

    From this analysis one can deduce {(see \cite{Gomes2020})} that there are asymptotic past and future vacuum states $\lvert n \rangle $ indistinguishable from the point of view of local observables but different in the global winding number  quantity $n$.  

    { A very important property of the 
    action obtained from \eqref{eq:lcolor} is its invariance} over the so called large gauge transformations.   These are gauge transformations not obtained directly from exponentiation of the Lie algebra variables i.e. they are not continuously connected to the identity. See \cite{Jackiw,Treiman:1986ep} for further study on those transformations.{  The large transformations relate thus two configurations of $A$ which can not be related by homotopy. Therefore} the winding numbers of the { field} configuration before and after the transformation must differ. On the vacuum states we may introduce this via a unitary operator    
    \begin{equation}
    \label{eq:largeGenerator}
    U_{1}\lvert n \rangle= \lvert n+1 \rangle,\  U_m=U^m_1\textrm{ and }U_{-1}=U^\dagger_1,
    \end{equation}
    that represents the action of a large gauge transformation that increases in 1 the winding number.

    In what regards the physical meaning, it is important to remember that even
    though the winding number by itself is a meaningless quantity, tunneling between sectors of different winding numbers do have physical relevance. In the path integral formalism it amounts to add to the Lagrangian \eqref{eq:lcolor} the CP violating term \cite{Weinberg1996}
    \begin{equation}
    \label{eq:thetatem}
    \mathcal{L}_{\theta}= \frac{\theta}{32 \pi^{2}} \operatorname{Tr}[ F \wedge {F}]
    \end{equation}
    which can be measured when coupled to massive fermions (quarks). Current measurements are compatible with $|\theta| \leq 10^{-10}$ raising the so called strong CP problem.

{
\subsection{$\theta$-vacua from locality and instantons} 
    
    If we assume that tunneling occurs in Nature we may study it via instanton configurations of the {E}uclidean action. Remember that instanton configurations are configurations of the gauge fields $A$ with nonvanishing Chern class and labelled by an integer which is often called the winding number of the instanton. All instantons with identical winding number are related by  gauge transformations belonging to the component of $SU(3)$ connected to the identity, also called small gauge transformations. By opposition, we will call large gauge transformations to those not connected with the identity. 
}
    Following \cite{Coleman1986} (Chap. 7 Sec 3.), we may assume that, because of the locality of \eqref{eq:chernterm}, an instanton of winding number $n$ on an Euclidean spacetime region  $\Omega_{[0,T]}$ that comprises a transition between times $0$ and $T$    can be decomposed, for sufficiently large $T$, in two instantons of winding numbers $n_1+n_2=n$ in disjoint regions $\Omega_{[0,T_1]}\cup\Omega_{[T_1,T]}$. From the  path integral perspective let the Euclidean action over the specified spacetime region be $S_{\Omega}=\int_{\Omega} d^4x \mathcal{L}_{E}(A)$ then we may compute the transition matrix

    \begin{equation}
    \label{eq:transition}
    F(T,n)=N \int_n [DA] \mathrm{e}^{-S_{\Omega_{[0,T]}}}
    \end{equation}
    where the subindex $n$ in the integral means that we must integrate over instanton configurations of winding number $n$. Then it follows that

    \begin{equation}
        \label{eq:transitiondesc}
        F(T,n)=\sum_{n_1+n_2=n}F(T_1,n_1)F(T-T_1,n_2).
        \end{equation}

   { If we take the Fourier transform $F(T,\theta)=\sum_{n}e^{-i\theta n}F(T,n)$, convolutions become} multiplications  $F(T,\theta)=F(T_1,\theta)F(T-T_1,\theta)$. Since $F(T,\theta)$ is to be interpreted as a transition matrix from an initial to a final state, and we have shown that { it} fulfills the same law as a time exponential, we may interpret it as a transition matrix of the evolution { of an eigenstate  $\lvert \theta \rangle$ of the Hamiltonian as 
   $F(T,\theta)  \propto\left\langle\theta\left|\mathrm{e}^{-H T}\right| \theta\right\rangle$, where}
\begin{equation}
\label{eq:thetavev}
    \begin{aligned}
        F(T, \theta)
        =N^{\prime} \int [DA] \mathrm{e}^{-S_{\Omega_{[0,T]}}} \mathrm{e}^{ \frac{i\theta}{8 \pi^{2}} \int_{\Omega_{[0,T]}}  \operatorname{Tr}[F\wedge F] }.
        \end{aligned}
    \end{equation}

{    The locality of operator \eqref{eq:chernterm} and assuming that tunneling exists,  leads  (see \cite{Coleman1986} ) to the $\theta$-vacuum 

    \begin{equation}
        \label{eq:thetavacuum}
        \lvert \theta \rangle=\sum_{n=-\infty}^\infty e^{-in\theta}\lvert n\rangle.
    \end{equation}

    This state is invariant under large gauge transformations because $U_{1}\lvert\theta\rangle = e^{i\theta} \lvert\theta\rangle$, i.e. they are proportional to the identity with a quantum mechanically irrelevant constant phase factor.

    Then in summary, if we consider only the gauge part of the theory,  the topological structure of the theta vacuum becomes a requirement derived from the existence of instanton solutions of non-trivial winding number together from the more fundamental condition of locality of operator \eqref{eq:chernterm}}. 


{
\section{The cluster decomposition principle (CDP) and confinement: Lowdon theorem}
    
\subsection{ $\theta$-vacuum implies CDP and vice versa} 

The Cluster Decomposition Principle (or property) uses locality of a theory to require the independence of measurements done in causally-disconnected regions of spacetime. This is equivalent to require the factorizability of expectation values of local operators on disconnected compact domains.
Following Weinberg (\cite{Weinberg1996}, section 23.6), if we consider a non-abelian Yang Mills theory in {E}uclidean space time and assume the existence of instanton configurations with different winding numbers $n$, then it follows that CDP holds} if and only if the vacuum of the theory is given by \eqref{eq:thetavacuum}. {
    Indeed, if} we compute  the expectation value of an observable $\mathcal{O}$ located within an {E}uclidean spacetime region $\Omega$ and an {E}uclidean Lagrangian density $\mathcal{L}(A)$ and action $S_{\Omega}=\int_\Omega d^4x \mathcal{L}(A)$, we obtain:
    \begin{equation}
    \label{eq:cluster}
    \langle \mathcal{O} \rangle_{\Omega}= \frac{\sum_{n}\omega(n) \int_n [D A] e^{-S_\Omega} \mathcal{O}(A)}{\sum_{n}\omega(n) \int_n [D A] e^{-S_\Omega}},
    \end{equation}
    where $\omega(n)$ are arbitrary  weight factors for each instanton configuration of winding number $n$. 

    Now if we let $\mathcal{O}$ to be located in a { smaller region $\Omega_1$ such that $\Omega_1 \cup \Omega_2= \Omega$ for $\Omega_1\cap\Omega_2=\emptyset$ then, because of the locallity of the winding number operator exploited in the previous section}, it follows 
    \begin{equation}
        \label{eq:cluster2}
        \langle \mathcal{O} \rangle_{\Omega}= \frac{
            \sum_{n,m}\omega(n+m)
            \int_n [D A] e^{-S_{\Omega_1}} \mathcal{O}(A)
            \int_m [D A] e^{-S_{\Omega_2}}
        }
        {
            \sum_{n,m}\omega(n+m)
            \int_n [D A] e^{-S_{\Omega_1}}
            \int_m [D A] e^{-S_{\Omega_2}}
        }.
    \end{equation}
    Now assume that the CDP holds, then region $\Omega_2$ should not contribute to the integral and therefore we must ensure $\langle \mathcal{O} \rangle_\Omega=\langle \mathcal{O} \rangle_{\Omega_1}$ which only happens if $\omega(n)=e^{in\theta}$ with $\theta$ an arbitrary parameter. Therefore, we conclude that from the existence of nontrivial configurations plus the Cluster decomposition principle it follows that the vacuum of QCD { must be  given by \eqref{eq:thetavacuum}.
    
    On the other hand if we take \eqref{eq:thetavacuum} as}  given then $\omega(n)=e^{-in\theta}$ and $\langle \mathcal{O} \rangle_\Omega=\langle \mathcal{O} \rangle_{\Omega_1}$ which is the statement of the CDP.
    This implies that the Cluster Decomposition Principle { can be} derived from the existence of instantons with non trivial winding number that, { as we saw in previous section, leads to the $\theta$--vacuum}. But it also implies that the $\theta$-vacuum is the { only vacuum} state compatible with the CDP if such instanton configurations are present.

    \subsection{Lowdon's theorem}
    
    Following \cite{Lowdon2016} we will assume no mass gap in the locally quantized linear space of QCD $\mathcal{V}$, i.e. no mass gap before implementing Becchi-Rouet-Stora-Tyutin  (BRST) reduction to the physical degrees of freedom of the theory. { From such a space, we construct the physical Hilbert space, which we will denote  as} $\mathcal{V}_{phys}$. With these assumptions,  Lowdon \cite{Lowdon2016} shows that the violation of the CDP implies that correlator strength between clusters of gluons increases at large distances{. This} is a sufficient condition for confinement. At this point it is worth remembering that the Standard Model of Elementary Particles has been constructed without resorting to the CDP.
    
    It  is important to notice that the absence of a mass gap in $\mathcal{V}$ does not exclude a mass gap in the physical space $\mathcal{V}_{phys}$. Nonetheless  there is no strong evidence that supports of disproves the existence of a mass gap in $\mathcal{V}$ while it is commonly accepted that $\mathcal{V}_{phys}$ does have a mass gap.   Under these assumptions certain lattice results {(negativity of the Schwinger functions for the quark and gluon propagators)} exposed in the aforementioned paper \cite{Lowdon2016} are understood as evidence for confinement.
 
    \section{Conclusion: screening tunneling implies confinement} 

    \subsection{A different structure of the vacua implies confinement}
    
     Let us now combine all the different properties discussed above.
    If tunneling between vacua with different winding numbers {was a} costly energetic process  {(this would mean a screened tunneling)} large gauge transformations $U_{n}$ would not be a symmetry of the theory. In this scenario we could exclude instanton configurations and therefore restrictions on the form of the $\theta$ vacuum would disappear.  Hence vacua different from   \eqref{eq:thetavacuum} would be possible, such as
    \begin{equation}
        \label{eq:thetavacuumgen}
        \lvert \omega \rangle=\sum_{n=-\infty}^\infty \omega(n)\lvert n\rangle,
    \end{equation}
    with more general weights $\omega(n)$.
    
    It is important to notice that these vacua would not be invariant under the so called large gauge transformations, i.e., we should exclude $U_{n}$  as a gauge symmetry leaving only the so called small gauge transformations (i.e., those connected to the identity) as a legitimate gauge transformation for the model. 
   In such a situation,  the restrictions on the topological structure of the vacuum  derived from  the CDP would no longer hold, and therefore  under the hypothesis made in the previous section, the vacuum may be different and this mechanism would lead to QCD confinement.

    We conclude that the change of the $\theta-$vacuum structure, for which the screening of tunneling (or, equivalently, the non-trivial topology of the gauge group) is required, would imply that CDP fails and this is a sufficient condition of confinement under the hypothesis made in \cite{Lowdon2016}.  
    
    \subsection{A mechanism to screen instantons that generates confinement}

    The final step in this discussion is to show a mechanism that screens instanton transitions, breaks the discrete global symmetry $U_n$ of \eqref{eq:largeGenerator}  and  produces a different vacuum. This is precisely the 
    kind of mechanism that  triggers the chain of implications exposed in this paper and produce confinement as discussed above.

    Our proposed mechanism is based on the axion formulation of \cite{dvali2022}. In this presentation an axion-like mechanism is written in { in the language of three forms} by promoting the Chern-Simons form \eqref{eq:ChernSimonsDensity}  to {become} a dynamical field $\operatorname{cs}_A$ {whose field-strength is} then {represented by} $d\operatorname{cs}_A$.  From this perspective this is a massless theory  with gauge symmetry and without propagating degrees of freedom. The mechanism is achieved by adding a two-form field $B$, that plays the dual role of the axion. In that system there is a Higgs phase in which the gauge redundant form $\operatorname{cs}_A$ acquires a mass by 'eating up' $B$. This axion mechanism leads to an {expectation} value of the Chern class \eqref{eq:chernterm} unambiguously zero effectively blocking tunneling between sectors of different winding number.

    Considering every contribution of the model besides gauge fields  the description is similar but slightly more involved. In principle we should study the QCD Lagrangian density plus the $\theta$ term \eqref{eq:thetatem} and every other possible contribution for our model.  Nonetheless, to explore the physics of the $\theta-$vacuum, we can argue,  studying the topological susceptibility of the vacuum  \cite{dvali2022},  that it is enough to study the theory that governs the massless modes of the 3-form field. 
    In an effective model we denote the massless modes of $\operatorname{cs}_A$ with an independent 3-form field $C=C_{\mu\nu\sigma}dx^{\mu}\wedge dx^{\nu}\wedge dx^{\sigma}$. We may interpret the coefficients of  $E=dC$, given by $E_{\rho\mu\nu\sigma}=\partial_{[\rho}C_{\mu\nu\sigma]}$, as the field stress tensor of the massless mode. It follows that this massless mode is described by the $\theta$-term of the Lagrangian  plus a kinetic term $\mathcal{K}(E)${, to be determined.
    T}hus we must study 
    \begin{equation}
    \label{eq:Lagrangian}
    L_\theta[C]=\mathcal{K}(E)+\frac{\theta}4E.
    \end{equation}
    Here $\mathcal{K}(E)$ must be an algebraic function of $E$ without linear terms. It is clear that this action presents a gauge symmetry 
    \begin{equation}
    \label{eq:gauge_invariance}
    C\to C+d \Theta 
    \end{equation}
    where $\Theta $ is any 2-form. Because of this symmetry, this field does not have propagating degrees of freedom and the solutions of the classical equations of motion are $
    E_{\rho\mu\nu\sigma}=E^0\epsilon_{\rho\mu\nu\sigma}
    $ where $E^0$ is an arbitrary constant referred to as the {\textit{electric}} field. We may interpret the $\theta$-term of the Lagrangian as a source term on the boundary and then fix $E^0=\theta$. From this {perspective} the different $\theta$-vacua are the superselection sectors of the theory corresponding to different values of $\theta$ or, analogously, different solutions of the classical equations of motion.

    The idea of the dualized axion mechanism of \cite{dvali2022} is to screen {the} {\textit{electric}} field $E$ by giving a mass to the 3-form field $C$ without breaking the gauge invariance. This way to proceed is equivalent to a regular axion mechanism with shift symmetry whose potential is completely determined by $\mathcal{K}$ (and vice versa). To do so we describe our effective picture in the Higgs phase adding to the Lagrangian the two form field $B$ coupled to $C$. Neglecting the $\theta$ boundary term this Lagrangian is written  as
    \begin{equation}
    \label{eq:axionlag}
    L[C,B]=\mathcal{K}(E)+m_a^2(C-dB)\wedge \ast (C-dB)
    \end{equation}   
    where $\ast$ is the Hodge operator.     
    This mass term is invariant under the symmetry 
    \begin{equation}
    \label{eq:symm}
        C\to C+d\Theta\text{ and }B\to B+\Theta
    \end{equation}
    and there is an additional gauge symmetry 
    \begin{equation}
        \label{eq:symmB}
           B\to B+d \xi
        \end{equation}
        where $\xi$ is an arbitrary 1-form. This implies { (see  \cite{dvaliThreeFormGaugingAxion2005})} that  $B$ propagates one degree of freedom and there are no massless modes for $C$. {Within} this mechanism  the instanton configurations are screened and replaced by classical solutions that decay exponentially fast and the electric field $E$ is unambiguously zero as well as the expected value of \eqref{eq:chernterm}.    
        In this phase quantum tunneling is mediated only by massive modes and thus $\lvert \langle n\vert m\rangle\rvert^2  \propto e^{-\alpha(\Omega) m_a^2\lvert n-m \rvert}$ where $\alpha(\Omega)$ is a constant that depends on the integration volume $\Omega$ that arises from the addition of the mass term $m_a^2 C_{\mu\nu\alpha}C^{\mu\nu\alpha}$ to the effective Lagrangian in \eqref{eq:transition}. This implies that {large \textit{gauge} transformations} are no longer a symmetry of the system.  {T}hus, in the Higgs phase, $\theta$-vacuum of this theory becomes  
        
        \begin{equation}
        \label{eq:axionvac}
        \lvert m_a\rangle=  \sum_{n=-\infty}^{\infty}  e^{-{m_a^2} \alpha(\Omega)  \lvert n\rvert-in\theta} \lvert n\rangle
        \end{equation}
        As we can see,  the term $\lvert n\rvert$ in the exponent spoils the equality between \eqref{eq:cluster} and \eqref{eq:cluster2} spoiling the CDP.    
        
        As a consequence of the axion-like mechanism described in this section, and under the hypothesis of \cite{Lowdon2016}, we conclude that confinement is derived as { a  result} of the violation of the CDP derived from the screening of tunneling between different sectors. { Indeed, any mechanism inducing} the same screening effect would be a sufficient condition for confinement in QCD.


\begin{acknowledgments}

\textbf{Acknowledgments}
The authors  would like to thank   A. Cherman for very useful discussions. Special mention deserves our thanks to Georgi Dvali for the correspondence on essential points of our work and his. The authors acknowledge partial financial support of Grant PID2021-123251NB-I00 funded by MCIN/AEI/10.13039/ 501100011033 and by the European Union, and of Grant E48-23R funded by Government of Aragon. C.B-M and D.M-C acknowledge financial support by Gobierno de Aragón through the grants defined in ORDEN IIU/1408/2018 and ORDEN CUS/581/2020 respectively. 
\end{acknowledgments}

\bibliography{QCDJL.bib}

\end{document}